\documentclass[12pt]{article}
\usepackage{graphicx}
\usepackage{color}
\usepackage{graphicx}
\input{color.pac}

\newsavebox{\sboxpubnumber}
\newsavebox{\sboxpubdate}
\newcommand{\pubdate}[1]{\begin{lrbox}{\sboxpubdate}{#1}\end{lrbox}}
\newcommand{\pubnumber}[1]{\begin{lrbox}{\sboxpubnumber}{\begin{tabular}{l} #1 \\
				 \usebox{\sboxpubdate}
				 \end{tabular}}
                           \end{lrbox}
                           \pubblock}
\newcommand{\Title}[1]{\begin{center} {\Large #1 } \end{center}}
\newcommand{\Author}[1]{\begin{center}{ \sc #1} \end{center}}
\newcommand{\Address}[1]{\begin{center}{ \it #1} \end{center}}
\newcommand{\andauth}{\begin{center}{and} \end{center}}
\newcommand{\pubblock}{\rightline{
			\usebox{\sboxpubnumber}}}
\newenvironment{Abstract}{\begin{quotation}  }{\end{quotation}}
\newenvironment{Presented}{\begin{quotation} \begin{center}
             PRESENTED AT\end{center}\bigskip
      \begin{center}\begin{large}}{\end{large}\end{center}
      \end{quotation}}

\newcommand{\crocho}{\left[\begin{array}{cc}}
\newcommand{\crochf}{\end{array}\right]}
\newcommand{\fb}{f_{\rm b}}
\newcommand{\fg}{f_{\rm gas}}
\newcommand{\h}{\:h_{50}}

\newcommand{\mb}{M_b}

\newcommand{\mprot}{m_{\rm p}}

\newcommand{\mst}{M_*}

\newcommand{\mtot}{M_{\rm tot}}
\newcommand{\omo}{\Omega_{\rm 0}}
\newcommand{\omM}{\Omega_{\rm m}}
\newcommand{\omb}{\Omega_{\rm b}}
\newcommand{\paro}{\left(\begin{array}{cc}\!\!}
\newcommand{\parf}{\!\!\end{array}\right)}

\newcommand{\rcx}{r_{\rm cX}}

\newcommand{\tx}{T_{\rm X}}

\begin{document}
%%%%%%%%%%%%%%%%%%%%%%%%%%%%%%%%%%%%%%%%%%%%%%%%%%%%%%%%%%%%%%%%%%%%%%%%
\begin{titlepage}
\pubdate{\today}                    %fill in the date
\pubnumber{XXX-XXXXX \\ YYY-YYYYYY} %preprint number(s)

\vfill
\Title{The baryon fraction in X-Rays galaxy clusters revisited.}
\vfill
\Author{R. Sadat}
\Address{LAT, Observatoire de Midi-Pyr\'en\'ees, 14 avenue Ed. Belin-
         31400- Toulouse}
\vfill
\andauth
\vfill
\Author{A. Blanchard}
\Address{LAT, Observatoire de Midi-Pyr\'en\'ees, 14 avenue Ed. Belin \\
         31400- Toulouse}
\andauth
\vfill
\Author{M. Douspis}
\Address{Astrophysics, Nuclear and Astrophysics Laboratory,
                   Keble Road,
		   Oxford, OX1 3RH,
	           UNITED KINGDOM}
\vfill
\begin{Abstract}
One of the most direct way to constrain the matter density of the universe $\Omega_M$ is to measure the baryon content in X-Rays clusters of galaxies. Typical value of the mean gas mass fraction is $\fg \sim 0.15h^{-3/2}$ which leads to $\omM <0.4$. In this talk I will discuss the issue of the gas fraction radial distribution inside a cluster and foccus on the apparent discrepancy between the theoretical shape predicted by numerical simulations and the observations. I will show that a significant part of this discrepancy is due to several systematics in both  the gas mass and total mass determinations. Revising the gas as well as the binding masses removes such a discrepancy and results in a lower baryon mass fraction of $\fb(r_{500}) \sim 10 \%$ (Sadat \& Blanchard 2001). Finally, I will discuss the cosmological constraints we obtain when we combine the revised baryon fraction with primordial fluctuations.
\end{Abstract}
\vfill
\begin{Presented}
    COSMO-01 \\
    Rovaniemi, Finland, \\
    August 29 -- September 4, 2001
\end{Presented}
\vfill
\end{titlepage}
\def\thefootnote{\fnsymbol{footnote}}
\setcounter{footnote}{0}

%%%%%%%%%%%%%%%%%%%%%%%%%%%%%%%%%%%%%%%%%%%%%%%%%%%%%%%%%%%%%%%%%%%%%%%%
% The document starts here
%%%%%%%%%%%%%%%%%%%%%%%%%%%%%%%%%%%%%%%%%%%%%%%%%%%%%%%%%%%%%%%%%%%%%%%%
\section{Introduction}

Clusters of galaxies are the most largest virialized systems in the Universe, hence the measurement of the baryonic to total mass ratio is representative of the cosmic value $\fb \equiv \frac{\omb^{bbn}}{\omM}$. The combination of the universal baryon density $\omb$ as predicted from light element abundances through the theory of big-bang nucleosynthesis with the cluster gas mass fraction, lead to a direct determination of $\omM$. 
This test to $\omM$ is in principle the most direct one as it depends on very few reasonable assumptions: the cluster formation by collapse from a well mixed medium, and no segregation between the gas and the dark matter as shown by numerical simulations.\\
In their pioneering paper, White and his collaborators have used Rosat observations of Coma cluster to estimate the gas mass fraction they found that the archetypical coma cluster has a baryonic content of $\mb \sim 3.4\h^{-5/2}10^{14}M_{\odot}$ with a very negligible contribution from stars $\mst \sim 2.2\h^{-1}10^{13}M_{\odot}$ and a binding mass of $\mtot \sim 2.2 \h^{-1}10^{15}M_{\odot}$. Hence they derived a gas fraction of $\fg \sim 0.15 \h^{-3/2}$ at Abell radius, which is several times higher than expected from the observed light element abundances ($0.024 \le \omb\h^{2} \le 0.08$) in $\omM=1$ universe, leading to the conclusion that we live in a low density universe (White et al. 1993). Several analyses of large samples of clusters have confirmed such a high baryon fraction with values ranging from $10-20\%$. These determinations depend on the radius where the gas mass fraction is estimated, since the gas fraction slightly increases with radius, but even when the gas fraction is determined in the same equivalent radius say $r_{500}$ (radius encompassing a mean density 500 times the critical density), current estimations are still discrepant (Evrard 1997). It has also be argued that the baryon fraction increases with cluster temperature (mass) (David et al. 1995, Mohr et al. 1999, Arnaud \& Evrard 1999). This has often been explained as due to non-gravitational processes such as heating effects occuring during galaxy cluster formation. In this talk I will address the issue of the baryon fraction in clusters by examining critically its distribution within the cluster and show how some systematic effects can bias the gas mass fraction determination. In the first section I will review the classical analysis method used to estimate gas mass fractions. Comparison between numerical simulations and observation is provided in section 2. Section 3 presents the systematic uncertainties on various steps involved in the gas mass fraction derivation and section 4 describes the revised gas mass fraction and the consequence on both the shape of the gas mass fraction profile as well as on the mean mass density of the universe. In section 5 we investigate the constraints on cosmological parameters by combining the cosmic microwave background (CMB) with the galaxy cluster baryon fraction.
\section{Gas mass fraction determination}
The standard way to derive the gas fraction proceeds from the following steps:\\
\noindent
$\bullet$ Assume a $\beta-model$ to fit the observed surface brightness $S_{o}(1+(\theta/\theta_{c})^2)^{-3\beta+1/2}$ where $\theta$ is the projected angular distance to the center. The gas density profile (hence the gas mass profile) is then derived using the two best-fit parameters $\beta$ and $\theta_{c}$.\\
$\bullet$ Use the hydrostatic equation to derive the total mass inside the same radius: $\mtot(r)=\frac{3k}{G \mu \mprot} \beta \tx r \paro 1 +
      \paro \frac{r}{\rcx} \parf ^{-2} \parf ^{-1}$.
The total mass thus depends linearly on both $\beta$ and $\tx$. Hence, if the slope of the gas density is poorly determined, it will have a drastic influence on the derived mass. 
 \\
\subsection{Shortcomings of this method}
The X-ray gas emission is rarely traced out to very large radii, so that gas fractions are measured only up to an X-rays limiting radius for which the signal-to-noise is good enough. Most of the time, it was necessary  to {\it extrapolate} to the virial radius $r_{200}$ or to some other outer radius like $r_{500}$. It is then essential to realize that gas mass is estimated at a radius at which the actual emission is poorly constrained, with a value of the order of the X-rays background or less. Moreover, the emissivity being dominated by the central region, a parametric fit will be rather insensitive to the outer part of the gas profile. Finally, gas masses are estimated assuming that the gas is smoothly distributed within the intracluster medium. Concerning total mass determination, the $\beta$ - model approach introduces an intrinsic scatter in the total mass estimates, of $\sim 30\%$ at $r_{500}$ which can be considerably reduced by using scaling relations between the mass and the temperature $T_{X}$ calibrated with numerical simulations (Evrard 1998).  Moreover, the $\beta$ - model approach leads to total masses which are systematically lower than what one would obtained using scaling relations between the mass and the temperature $T_{X}$ calibrated with numerical simulations (Roussel et al. 2000, hereafter RSB00). However, some problems exist in this latter approach. There is a scatter in the $M -T_{X}$ relation existing among different authors: 
$\tx = 4.75 (M_{200}/10^{15}M_{\odot})^{2/3} \textnormal{keV}$ (Evrard, Metzler \& Navarro 1996, hereafter EMN96) and 
$\tx = 3.81 (M_{178}/10^{15}M_{\odot})^{2/3}$ \textnormal{keV} (Bryan and Norman 1998, BN98); these two normalizations being the most extreme values. There is also the problem that most of the numerical simulations do not include the gas cooling neither the heating of the gas with supernovae or AGN, which may introduce a further change in the relation, although these effects might play minor role in the case of hot and massive clusters.\\
\section{The baryon fraction profile: Observations versus simulations}
\begin{figure}[htb]
    \centering
    \includegraphics[height=3.5in]{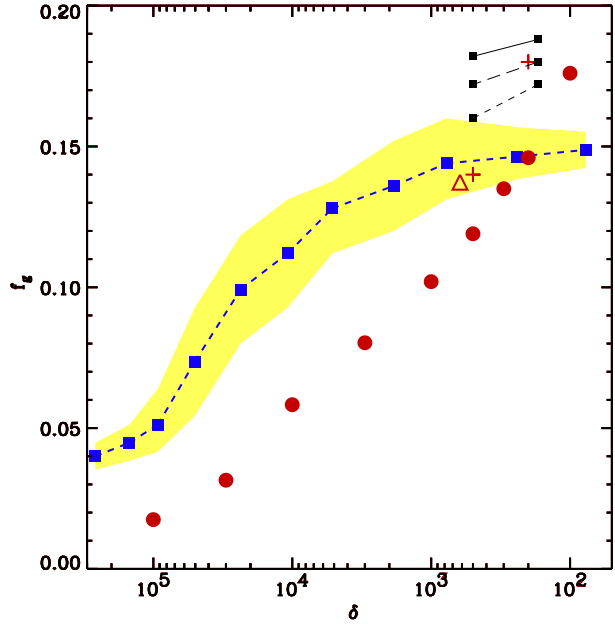}
    \caption{The distribution of the gas fraction $f_g$ versus $\delta$ ($\delta \equiv \rho(<r)/\rho_c$ where 
$\rho_c \equiv 3H^{2}_{o}/8\pi G$). The observations are from RSB00 
 {\it filled circles}, Ettori\&Fabian 1999 {\it empty triangle} and Arnaud \& Evrard 1999 {\it crosses}. Statistical uncertainties on these quantities are smaller than the symbols size. The {\it squares} connected by a 
dashed line correspond to the theoretical gas fraction for a global value of $16\%$. {\it Small squares} connected by a line correspond to the gas fraction calculated from 
numerical simulations including winds by Metzler \& Evrard (1997): for a 6 keV cluster ({\it long dashes}), a 3 keV cluster ({\it small dashes}) and without including winds ({\it continuous line}) with a global value of 20\%.}
    \label{fig:fbobs}
\end{figure}

In the case where only gravity is acting during cluster formation, the gas fraction profile is expected to follow a scaling law, depending only on the contrast density $\delta$. Hydro-dynamical simulations of an X-rays cluster from different groups (the Santa Barbara group) have shown that the gas fraction normalized to the global value, increases in the inner part and then tends to flatten in the outer part to reach the cosmic value see Fig. \ref{fig:fbobs}. On the observational side, the distribution of the gas fraction within clusters has been widely examined by RSB00 for a large sample of groups and galaxy clusters. The resulting shape is shown in (Fig. \ref{fig:fbobs}). To ensure consistency gas fractions - calibrated by the numerical simulation - published in RSB00 are used. For comparison gas fractions at $r_{500}$ and $r_{200}$ from other authors are also plotted in (Fig. \ref{fig:fbobs}). The comparison of both profiles is very surprising as it shows that:
the gas fraction profile derived from observations is in strong disagreement with the numerical simulations results with a global value of $16\%$, its shape continuously increases and does not exhibit the flattening in the outer parts as seen in numerical simulations.
Clearly, this discrepancy calls for caution when one is using the gas fraction to set upper limit on the mean density of the universe.\\
One possibility is that processes acting during the cluster formation are not well understood and probably non-gravitational processes could have played an important role. But this is probably not the case, because large departure from scaling laws is expected which is not observed in the data (RSB00).

%%%%%%%%%%%%%%%%%%%%%%%%%%%%%%%%%%%%%%%%%%%%%%%%%%%%%%%%%%%%%%%%%%%%%%%%
%%
%%   use this format to include an .eps figure into your paper
%%
%%%%%%%%%%%%%%%%%%%%%%%%%%%%%%%%%%%%%%%%%%%%%%%%%%%%%%%%%%%%%%%%%%%%%%%%
\section{Uncertainties in gas mass estimation}
\subsection{The extrapolation problem}
X-rays gas masses are usually derived up to an X-rays limiting radius $R_{Xlim}$ where the signal-to-noise is good enough. Most of the time it was necessay to extrapolate the $\beta$-model up to the virial radius $r_{200}$, where the emission is actually dominated by the background.  Moreover a parametric fit will be rather insensitive to the outer part, because the emissivity is dominated by the central region. This crude extrapolation may then raise some question on its validity. 
In a recent study of a large sample of X-rays ROSAT images in which the data trace the emission up to very large radii (approximately up to the virial radius), it has been shown that the $\beta$-model does not provide an accurate description of the surface brightness over the whole range of radii. They have found that the outer slopes (0.3$r_{200}$ to $r_{200}$) are actually steeper than the slope found when the $\beta$-model is applied to the whole cluster. Hence, their derived gas masses are lower than what one would obtain by extrapolating the $\beta$-model. 
\subsection{The clumping effect}
In most of the studies, the ICM gas is assumed to be uniform, which is not true, as actual clusters do exhibit a certain level of density fluctuations at small and large scale due probably to accretion and mergers events. This clumping is assumed to be more pronouced in the outer parts of the clusters where the relaxation is not completely achieved. In the presence of clumping the gas masses are overestimated by a factor $\sqrt{C}$ where $C = <\rho_g^2>/<\rho_g>^2$ ($\rho_g$ is the gas density). This factor has been quantified from numerical simulations by Mathiesen et al. (1999). They found that models assuming a uniform density introduce a bias of $\sqrt{C} \sim 1.16$ at $\delta = 500$. 

\section{The mean gas fraction revisited: consequence on gas fraction profile and $\omM$}
\begin{figure}[htb]
    \centering
    \includegraphics[height=3.5in]{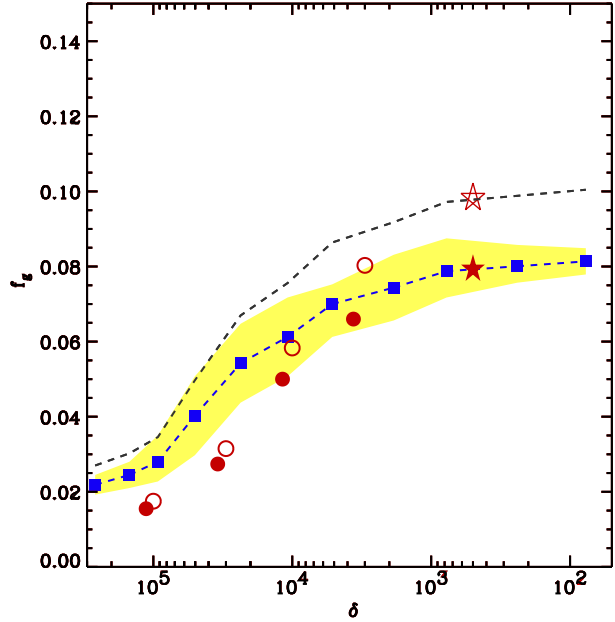}
    \caption{The distribution of the gas fraction $f_g$ versus $\delta$: {\it circles} correspond to the observed $f_g$ from RSB00 but restricted to the region where the gas is detected (no extrapolation), using EMN96 ({\it empty circles}) and BN98 ({\it filled circles}) calibration. {\it  Empty} ({\it filled}) {\it star} corresponds to the estimated $f_g$ using VFJ99 sample and corrected for the clumping with EMN96 (BN98) calibration. {\it Filled squares} linked by a dashed line correspond to the gas fraction predicted by numerical simulations for a global fraction of 8.75\%. The upper dashed line corresponds to the case where the global fraction is 10.8\%}
    \label{fig:fb_res}
\end{figure}

Here, the gas masses are corrected for the extrapolation problem by using the Vikhlinin et al. (1999) sample which is the only sample for which the emissivity is detected up to virial regions and then corrected the gas mass at $\delta=500$ for the clumping effect using the Mathiesen et al. (1999) factor.  The total mass is derived from scaling relations calibrated from numerical simulations (using both EMN96 and BN98 calibrations). The result is plotted in figure \ref{fig:fb_res}. From this figure, we can see that the gas fraction distribution, in particular the asymptotic behaviour, is now consistent with numerical simulations predictions, (although a slight difference remains in the inner parts at $\delta <10^{4}$ which can well be due to energy injection, not negligible in these regions). The resulting gas fraction at $r_{500}$ is now lower than previous estimates. By matching the data point at $\delta=500$, we find $\fg=0.0875 \pm 0.0075$ ($0.108\pm0.0092$) at $68\%$ confidence level with BN98 (EMN96) normalization, the uncertainty being due to numerical simulations (Fig. \ref{fig:fb_res}). Adding the stellar mean contribution of $1\%$, our final value of the baryon fraction at the virial radius is:  $\fb \sim 0.10 \pm 0.01$ (with BN98 calibration) and $\fb \sim 0.12 \pm 0.01$ (with EMN96 calibration). Combining the derived baryon fraction with the universal baryon density $\omb \sim 0.08$ as derived from recent D/H measurement reported by O'Meara et al. 2000 (this value has been confirmed by recent results on CMB fluctuations measurements from Boomerang mission 2001), we find the following value for $\omo$:
$\omo = 0.8 \pm 0.1$ using the baryon fraction estimated with BN98 normalization.
%%%%%%%%%%%%%%%%%%%%%%%%%%%%%%%%%%%%%%%%%%%%%%%%%%%%%%%%%%%%%%%%%%%%%%%%
%%
%%   use this format to include an .eps figure into your paper
%%
%%%%%%%%%%%%%%%%%%%%%%%%%%%%%%%%%%%%%%%%%%%%%%%%%%%%%%%%%%%%%%%%%%%%%%%%
%%
%%   use this format to include an .eps figure into your paper
%%
\begin{figure}[htb]
    \centering
    \includegraphics[height=3.5in]{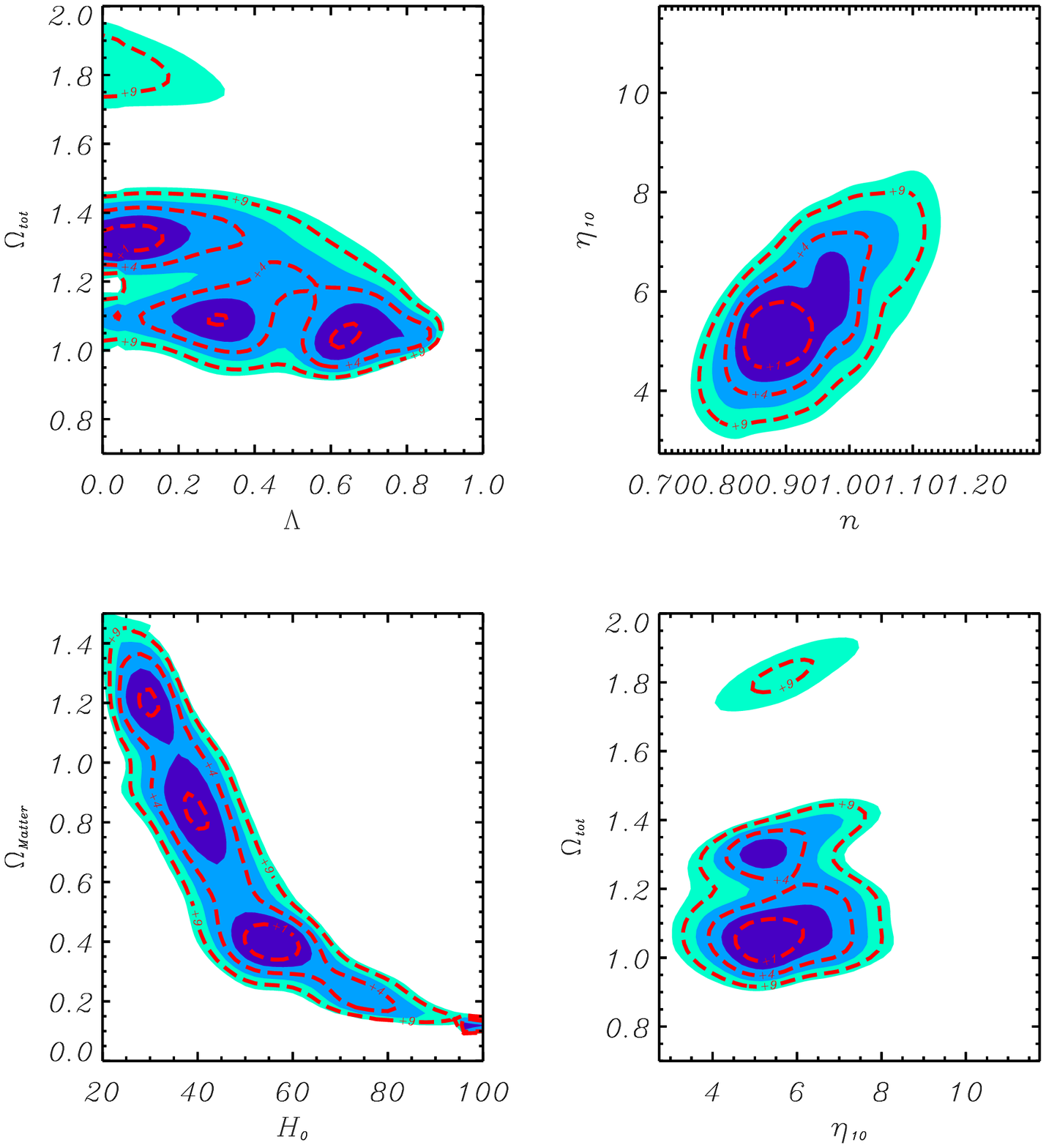}
    \caption{Different contour plots from the analysis of the most recent CMB data (COBE, BOOMERANG, MAXIMA, DASI). The dashed red line define the 68, 95 and 99\% GCL when
projected on the axis. The blue filled contours are the corresponding
confidence intervals in 2 dimensions.}
    \label{fig:fig_cmb}
\end{figure}

%%%%%%%%%%%%%%%%%%%%%%%%%%%%%%%%%%%%%%%%%%%%%%%%%%%%%%%%%%%%%%%%%%%%%%%%
%%%%%%%%%%%%%%%%%%%%%%%%%%%%%%%%%%%%%%%%%%%%%%%%%%%%%%%%%%%%%%%%%%%%%%%%

\section{Combining CMB with the baryon fraction: new constraints on the cosmological parameters}

%%%%%%%%%%%%%%%%%%%%%%%%%%%%%%%%%%%%%%%%%%%%%%%%%%%%%%%%%%%%%%%%%%%%%%%%

%%
%%   use this format to include an .eps figure into your paper
%%
\begin{figure*}[h]
	\begin{minipage}[t]{7.8cm}
    \includegraphics[height=3.5in]{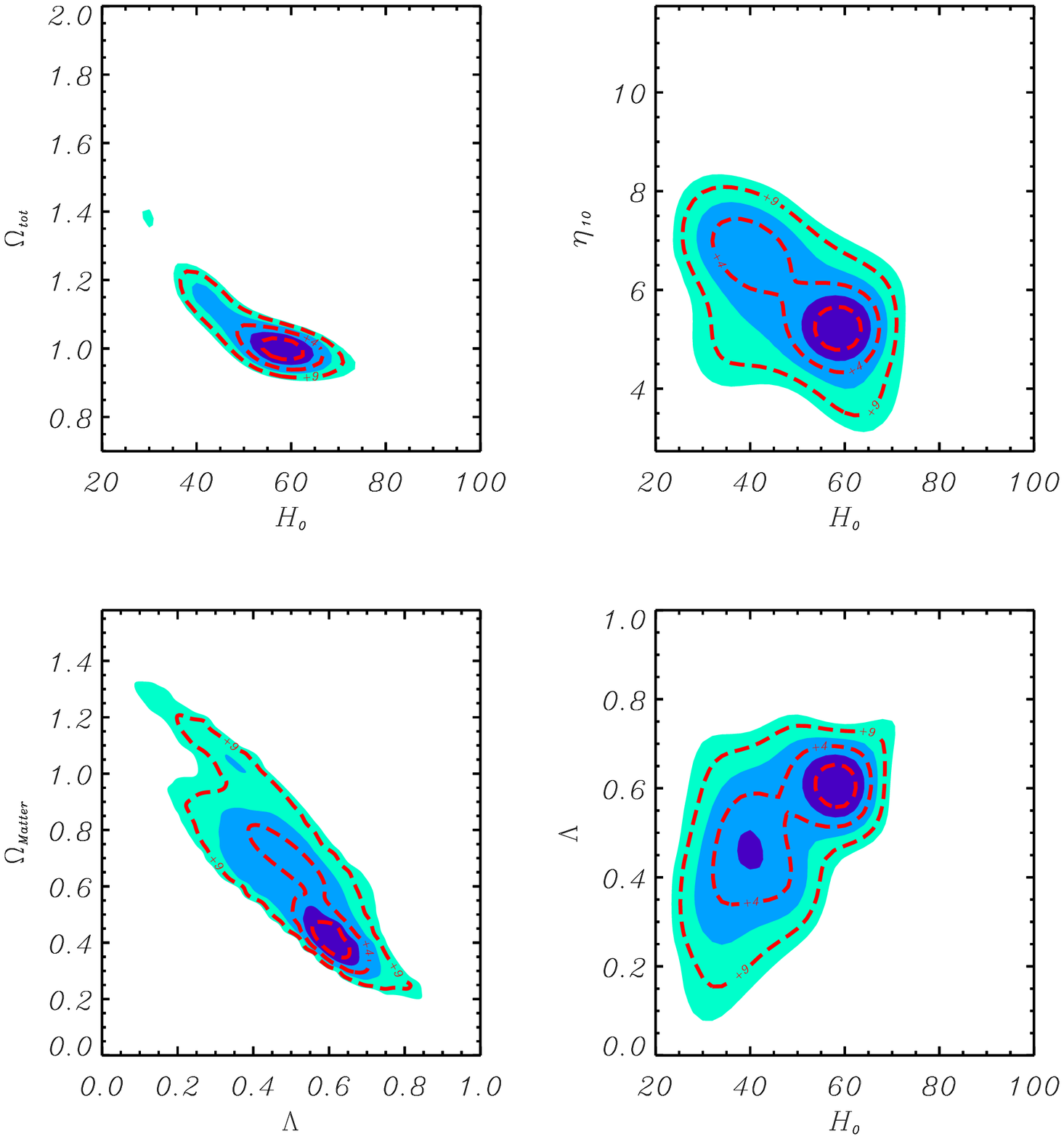}
	\end{minipage}
	\begin{minipage}[t]{7.8cm}
	\includegraphics[height=3.5in]{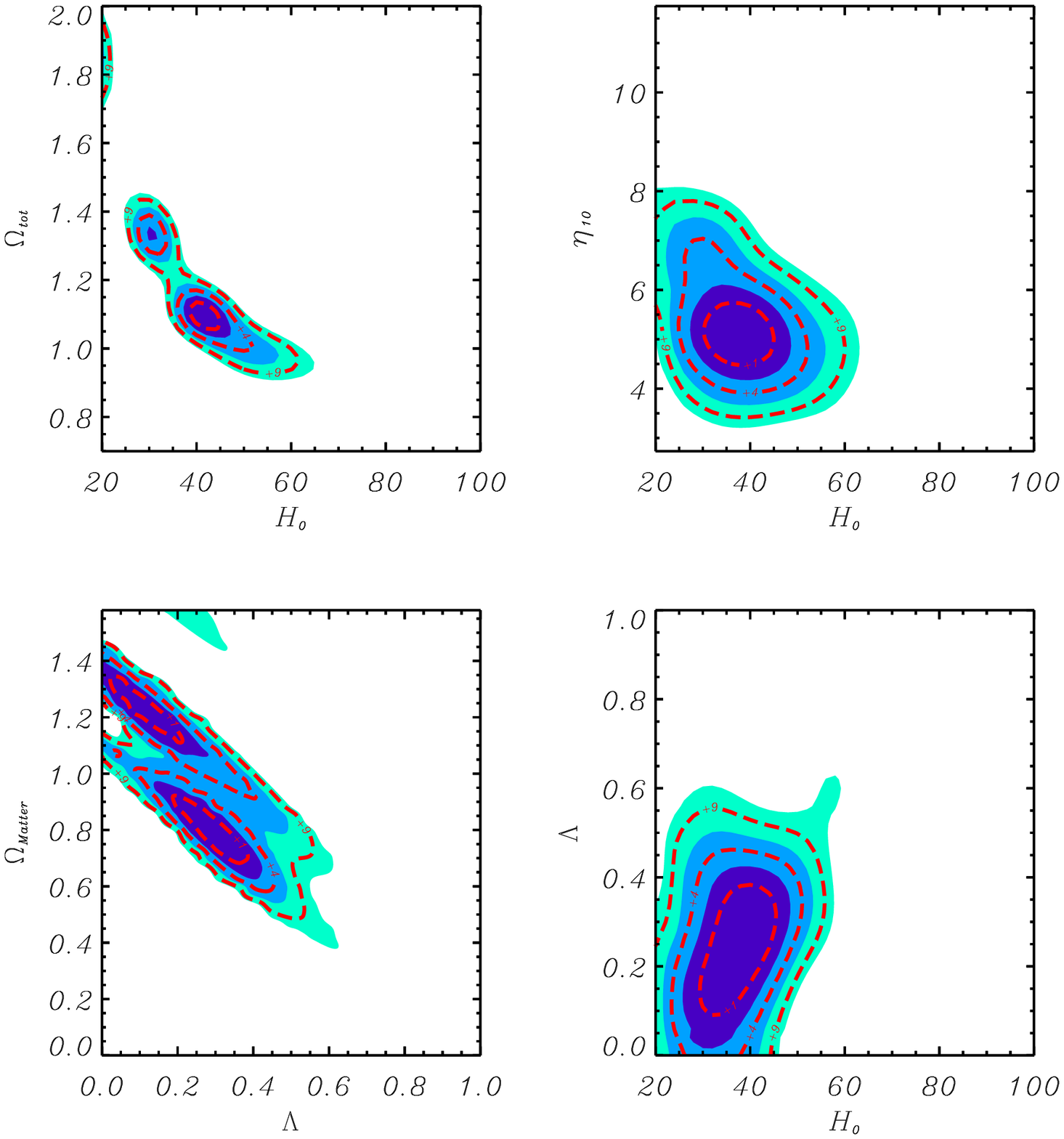}
	\end{minipage}
    \caption{Combined analysis of CMB and the baryon fraction. {\it Left}: the high baryon fraction value case with $\fb \approx 15\%$, {\it right}: the low baryon fraction case with $\fb \approx 10\%$. The definition of contours is same as in Fig. 3}
    \label{fig:fig_cmbhfb}
\end{figure*}

%%%%%%%%%%%%%%%%%%%%%%%%%%%%%%%%%%%%%%%%%%%%%%%%%%%%%%%%%%%%%%%%%%%%%%%% 
Given the fact that CMB implies degeneracies between the fundamental
cosmological parameters $H_0$, $ \Omega_\Lambda$
and $\Omega_m$, (see Fig. 3) it is interesting to look for combination with constraint that does not measure directly any of these parameters. Here I will present the results we obtain when we investigate the combination with the baryon fraction inferred from cluster observations using both the {\it high baryon fraction} as found in the literature and the revised {\it low baryon fraction} inferred in this work (Douspis et al. 2001).

\subsubsection{The high baryon fraction case}

The main result from adding the constraint from the baryon fraction is to restrict significantly the various contours obtained with CMB alone (see figure \ref{fig_cmbhfb}). This is very clear from
the $H_0-\eta_{10}$ plane: the area of the contours is much
reduced, leading to a preferred model which is around $H_0 \sim 60 $
km/s/Mpc and $\eta_{10}\sim 5$. The other two-dimensional plots also
reveal that  all the contours are significantly reduced: actually most
of the one sigma contours are of the order of the grid size so that
the best model and 1$\sigma$ ranges should be used with caution. It is
highly interesting that the cosmological model is now very well
constrained: 
$\Omega_{\lambda} \sim 0.6$,   $\Omega_{tot} \sim 1.$ (implying
$\Omega_{m} \sim 0.4$,  $H_0 \sim 60 $ km/s/Mpc, $\eta_{10}\sim 5$,
$n \sim 0.85$). As we mentioned the allowed range have to be
interpreted with caution,  yet it is worth noticing that the preferred value of
the Hubble constant is rather low and that higher values like
80 km/s/Mpc  lie uncomfortably outside of the preferred region.

\subsubsection{The low baryon fraction case}

Using the low value for the baryon fraction leads similarly to much restricted
contours. However the latter  differ from the contours obtained with a higher baryon fraction : the 
preferred value for $\eta_{10}$ is again close to 5., the preferred value of 
the Hubble constant is now very low ($H_0 \sim 40 $ km/s/Mpc). It is worth
 noticing that values as $H_0 \sim 60 $ km/s/Mpc are still at the
edge of the 3  
sigma  contour and for this reason cannot be entirely ruled out. The preferred
 model is $\Omega_{\lambda} \sim 0.3$,   $\Omega_{tot} \sim 1.1$. This implies an $\Omega_{m} \sim 0.8$ with $H_0 \sim 40 $ km/s/Mpc, $\eta_{10}\sim 5.$ and $n \sim 0.85$ (see figure 4). 

\section{Conclusions}
The baryon fraction in X-rays clusters is one of the most direct way to constrain the density of the universe. Previous estimations of the baryon fraction yield to a high value $\fb \sim 15\% \h^{-3/2}$, favouring low density universe $\omo \sim 0.3$. In my talk I have shown that the radial profil of the gas fraction as derived from observations differs strongly from numerical simulations results. I argue this is not due to non-gravitational processes which may occur during the cluster formation but is rather due to various systematics such as the extrapolation of a $\beta$-model and the clumpiness of the gas that have not been taken into account in previous estimations. Correcting the gas masses from these biases and using total masses calibrated from numerical simulations yield to a better agreement between observations and numerical simulations for a global value of the order of $\sim 10 (12)\%\h^{-3/2}$ depending on which normalization we use to compute the total mass. This revised value is lower than previous ones and leads to $\omo \sim 0.8 (0.66)\h^{-1/2}$. \\
Finally combining the most recent CMB data with the baryon fraction leads to very interesting constraints on the various parameters. Using the revised value ({\it the low baryon fraction value}), we found that the best model is
$\Omega_{\lambda} \sim 0.3$,   $\Omega_{tot} \sim 1.1$ ,  $H_0 \sim 40
$ km/s/Mpc, $\eta_{10}\sim 5.$, $n \sim 0.85$ implying a high density
parameter  $\Omega_{m} \sim 0.8$ consistent with the determination
from the baryon fraction (Sadat \& Blanchard 2001) and cluster abundance evolution (Blanchard et al. 2000).


\begin{thebibliography}{99}
\bibitem{Arnaud}{}{}
Arnaud, M. \& Evrard, A.E., MNRAS {305}, {631} (1999)

\bibitem{Blanchard2}{}{}
Blanchard, A., Sadat, R.,   Bartlett, J. \& Le Dour, M., A\&A {362}, {809} (2000)

\bibitem{Bryan}{}{}
Bryan, G.L. \& Norman, M.L., ApJ. {495}, {80} (1998) 

\bibitem{David et al.}{}{}
David, L.P., Jones, C. \& Forman, W., ApJ. {445},{578} (1995)

\bibitem{Douspis et al.}{}{}
Douspis, M ., Blanchard, A., Sadat, R., Bartlett, J.G., Le Dour, M, A\&A {379}, {1D} (2001)

\bibitem{Ettori}{}{}
Ettori, S. \& Fabian, A.C., MNRAS {305},{834} (1999)

\bibitem{Evrard}{}{}
Evrard, A.E, MNRAS {292}, {289} (1997)

\bibitem{Evrard et al.}{}{}
Evrard, A.E, Metzler, C.A., \& Navarro, J.F., ApJ. {469}, {494} (1996)

\bibitem{{Frenk} et al.}{}{}
Frenk, C.S, White, S.D.M et al., ApJ. {525}, {554} (1999)  


\bibitem{Mathiesen et al.}{}{}
Mathiesen, B., Evrard, A.E. \& Mohr, J.J., ApJ. {520}, {L21} (1999)

\bibitem{Metzler}{}{}
Metzler, C.A. \&  Evrard, A.E., astro-ph/9710324 (1997)

\bibitem{Mathiesen et al.}{}{}
Mohr, J.J.,Mathiesen, B., and Evrard, A.E, ApJ. {517}, {627} (1999)

\bibitem {}Navarro J.\ F., Frenk C.\ S. \& White S.\ D.\ M., MNRAS {275}, {720} (1995)(NFW)


\bibitem{O'Meara et al.}{}{}O'Meara, J.M. et al., AAS {197}, {5604} (2000)

\bibitem{Roussel et al.}
Roussel, H., Sadat, R. \& Blanchard, A., A\&A {361}{429}, (2000)

\bibitem{Sadat et al.}{}{}
Sadat, R. \& Blanchard, A., A\&A {371}, {19} (2001)

\bibitem{Tegmark.}{}{}
Tegmark M. \& Hamilton A. 1997, astro--ph/9702019

\bibitem{Vikhlinin et al.}{}{}
{Vikhlinin}, A., {Forman}, W. and {Jones}, C., ApJ., {525}, {47} (1999)


\bibitem{White et al.}{}{}
White, S.D.M., Navarro, J.F., Evrard, A.E. \& Frenk, C., Nat. {366}, {429} (1993)


\end{thebibliography}
\end{document}